\documentclass[12pt,a4paper,final]{iopart}
\usepackage{iopams}  
\usepackage{graphicx}
\usepackage[breaklinks=true,colorlinks=true,linkcolor=blue,urlcolor=blue,citecolor=blue]{hyperref}
\usepackage{soul}

\begin{document}

\title[Electrostatic sorter for electrons carrying orbital angular momentum]{Design of electrostatic phase elements for sorting the orbital angular momentum of electrons}

\author{Giulio Pozzi$^{1,2}$}
\address{$^1$Ernst Ruska-Centre for Microscopy and Spectroscopy with Electrons 
and  Peter Gr\"unberg Institute, Forschungzentrum~J\"ulich,~52425~J\"ulich,~Germany}
\address{$^2$Department of  Physics and Astronomy, University of Bologna, viale B. Pichat 6/2, 40127 Bologna, Italy}
\ead{giulio.pozzi@unibo.it}

\author[cor1]{Vincenzo Grillo}
\address{CNR-Institute of Nanoscience-S3, via G. Campi 213/a, 41125 Modena, Italy}
\ead{vincenzo.grillo@unimore.it}

\author{Peng-Han Lu}
\address{Ernst Ruska-Centre for Microscopy and Spectroscopy with Electrons 
and  Peter Gr\"unberg Institute, Forschungzentrum~J\"ulich,~52425~J\"ulich,~Germany}
\ead{p.lu@fz-juelich.de}

\author{Amir H. Tavabi}
\address{Ernst Ruska-Centre for Microscopy and Spectroscopy with Electrons 
and  Peter Gr\"unberg Institute, Forschungzentrum~J\"ulich,~52425~J\"ulich,~Germany}
\ead{a.tavabi@fz-juelich.de}

\author{Ebrahim~Karimi}
\address{Department of Physics, University of Ottawa, 25 Templeton Street, Ottawa, Ontario K1N 6N5, Canada}
\ead{ekarimi@uottawa.ca}

\author{Rafal  E. Dunin-Borkowski}
\address{Ernst Ruska-Centre for Microscopy and Spectroscopy with Electrons 
and  Peter Gr\"unberg Institute, Forschungzentrum~J\"ulich,~52425~J\"ulich,~Germany}
\ead{r.dunin-borkowski@fz-juelich.de}

\begin{abstract}
The orbital angular momentum (OAM) sorter is  a new  electron optical  device  for measuring an electron's OAM. It is based on two phase elements, which are referred to as the \textquotedblleft unwrapper\textquotedblright and \textquotedblleft corrector\textquotedblright and are placed in Fourier-conjugate planes in an electron microscope. The most convenient implementation of this concept is based on the use of electrostatic phase elements, such as a charged needle as the unwrapper and a set of electrodes with alternating charges as the corrector. Here, we use simulations to assess the role of imperfections in such a device, in comparison to an ideal sorter. We show that the finite length of the needle and the boundary conditions introduce astigmatism, which leads to detrimental cross-talk in the OAM spectrum. We demonstrate that an improved setup comprising three charged needles can be used to compensate for this aberration, allowing measurements with a level of cross-talk in the OAM spectrum that is comparable to the ideal case.
\end{abstract}
\vspace{2pc}

\section{Introduction}

In quantum mechanics, two quantities -- momentum and angular momentum -- are linked to fundamental symmetries of space \cite{Noether:1918,Neuenschwander:2017}. The measurement of both quantities allows the translational and rotational symmetry of a system to be defined. In transmission electron microscopy (TEM), the momentum of an electron is a well-known quantity. One of the characteristic working modes of a TEM -- diffraction -- records a spectrum of the in-plane components of the electron's momentum. In contrast, the \emph{angular} momentum of the electron has only recently attracted interest, in particular through the study of its eigenstates, \emph{i.e.}, electron vortex beams \cite{Bliokh:2007,Uchida:2010,Verbeeck:2010,Mcmorran:2011,Harris:2015,Mafakheri:2017}. In order to make the angular momentum (and in particular its orbital component) a useful quantity, a new measurement r\'{e}gime is required that makes the OAM readily measurable in the form of a spectrum, just as for diffraction.

The ability to measure the OAM states of electrons in the TEM would be valuable not only in materials science (\emph{e.g.}, for measuring magnetic dichroism \cite{Verbeeck:2010}) but also for fundamental physics experiments that make use of OAM as a discrete spectrum variable for freely propagating electrons \cite{Harris:2015,Bliokh:2017,Larocque:2018}. Two recent studies~\cite{Grillo:2017,Mcmorran:2017} have made use of experiments and simulations to demonstrate that a set of electron optical phase elements, whose combination is referred to as a \textquotedblleft sorter\textquotedblright, can be used to sort the OAM states of electrons, based on an equivalent concept in light optics \cite{Berkhout:2010,Lavery:2012}. Just as in light optics, two phase elements are placed in different planes (the image and focal planes of a lens), in order to produce a conformal mapping of the wavefunction \cite{Bryngdahl:1974a,Bryngdahl:1974b,Saito:1983,Cederquist:1984,Hossack:1987} that transforms Cartesian to log-polar coordinates, thereby allowing an OAM spectrum to be extracted from its Fourier transform, \emph{i.e.}, using diffraction.

The technique is based on the application of a controlled phase distribution to the wavefunction of the incoming electron beam, typically using either diffractive or refractive optical elements. It is therefore necessary to have efficient phase elements that can be used to manipulate the electron wavefunction. One approach involves the introduction of phase holograms that are made from nanofabricated SiN films of varying thickness \cite{Grillo:2017,Grillo:2014}. However, such an approach results in a reduction in efficiency and does not allow the phase to be tuned dynamically. In light optics, the phase of a light beam can be adjusted using a spatial light modulator. Despite recent proposals \cite{Verbeeck:2018}, electron microscopes cannot yet be equipped with devices that are able to impart arbitrary phase distributions to electrons. An important example of progress towards achieving adaptive optics for electrons, albeit without full flexibility, is provided by the development of hardware for correcting the spherical aberration of electron lenses, most powerfully by using magnetic multipole elements to compensate for different orders of aberrations \cite{Haider:1998} and more simply by using thin film phase masks \cite{Grillo:2017a,Linck:2017,Shiloh:2018}.

Recent effort has been made to develop structured electrostatic fields that can be used as tuneable phase elements for generating and sorting electron OAM states \cite{Mcmorran:2017}. The phase distributions of such electrostatic arrangements had previously been studied numerically, analytically and experimentally \cite{Beleggia:2014,Frabboni:1987,Beleggia:2000a,Pozzi:2016}. However, their applications to electrostatic OAM sorters were not then explored. Here, we present simulations and discussions about the design and practical implementation of electrostatic phase elements as OAM sorters. We show that deviations in their phase distributions from the ideal situation result primarily from astigmatism associated with the finite length of a charged wire and boundary conditions. We propose a new device formed from three charged needles, which compensates for this aberration and thereby strongly reduces cross-talk in a sorted OAM spectrum.

\section{The ideal sorter}

The key optical component of an ideal sorter transforms the azimuthal coordinate in an input beam into a linear transverse coordinate in an output beam. An input image comprising concentric circles is then transformed into an output image comprising parallel lines. However, this transformation introduces a phase distortion that needs to be corrected by a second element, which is positioned in the Fourier plane of the first element. The complete system therefore comprises two elements: one to transform the coordinates and the other to correct for the phase distortion \cite{Berkhout:2010}.

The phase profile of the transforming optical element, which is referred to as the \textquotedblleft unwrapper\textquotedblright or \textquotedblleft sorter~1\textquotedblright, is given by the expression
\begin{equation}
\phi_1(x,y)=\frac{ d}{\lambda f} \left[ y \,\arctan{\left(\frac{y}{x}\right)}- x\ln \left(\frac{\sqrt{x^2+y^2}}{b} \right)+x\right], \label{optsort}
\end{equation}
which performs the conformal mapping $(x ,y)\rightarrow (u, v)$ in its Fourier plane $(u, v)$, where 
\begin{equation}
u=-\frac{d}{2\pi} \ln \left(\frac{\sqrt{x^2+y^2}}{b} \right),
\end{equation}
\begin{equation}
v= \frac{d}{ 2 \pi} \arctan{\left(\frac{y}{x}\right)},
\end{equation}
\textquotedblleft ln\textquotedblright is the Napierian logarithm and $\arctan{\left(\frac{y}{x}\right)}$ is defined as the angle in the Euclidean plane between the positive $x$~axis and the ray to the point $(x,y)$. In Eq.~\ref{optsort}, $\lambda $ is the wavelength of the incoming electron beam and $f$ is the focal length of the Fourier transforming lens. The parameter $d$ is  the length of the transformed beam, while the parameter $b$ translates the transformed image in the $u$ direction and can be chosen independently of $d$~\cite{Berkhout:2010}. It is important to note that the ideal transforming element contains a line of discontinuity, whose end defines the axis about which the OAM is measured.

The phase correction is given by a second element, which is referred to as the \textquotedblleft corrector\textquotedblright or \textquotedblleft sorter~2\textquotedblright and is defined by the expression,
\begin{equation} \label{optcorr}
\phi_2(u,v)=- \frac{d b}{\lambda f}  \exp \left(-  2 \pi \frac{u}{d} \right) \cos  \left(- 2 \pi \frac{v}{d} \right).
\end{equation}
In the optical case, if refractive elements are used instead of diffractive spatial light modulators, it has been found convenient to add to $\phi_1$ and $\phi_2$ a  thin convergent spherical lens contribution, of focal length $f$ equal to the  distance between the two elements, with the transmission function
\begin{equation}
\psi_{f}(x,y)=\exp \left[ \frac{-i \pi (x^2+y^2)}{\lambda f} \right],
\end{equation}
in order to have a very efficient and compact  mode transformer \cite{Lavery:2012}.

This setup can be adapted to the electron optical case. The sorter elements are therefore finally characterised by the transmission functions 
\begin{equation} \label{S1piuf}
 \exp[ i \phi_1(x,y)] \psi_{f}(x,y)
\end{equation}
and 
\begin{equation} \label{S2piuf}
\exp[i \phi_2(u,v)] \psi_{f}(u,v)
\end{equation}
for sorter~1 and sorter~2, respectively.

If a second Fourier transforming lens of focal length $F$ is added after the  phase correcting element, then the OAM states can be separated in the focal plane of this lens, provided that the stationary phase approximation is satisfied, \emph{i.e.}, that the phase variation of the beam is much lower than that of the sorter. This requirement can be satisfied if, at a given radius $r=R$,  the average phase variation $\theta$ per azimuth of the sorter is greater than the azimuthal gradient of the beam phase $\phi_{beam}(r,\theta)$ to be sorted, according to the expression
\begin{equation}
\frac{\phi_{1}(r=R,\theta=\pi)-\phi_{1}(r=R,\theta=0) }{2 \pi}\gg \nabla_{\theta}\, \phi_{beam}(r,\theta).
\end{equation}
The phase variation of the sorter is given approximately by the relation
\begin{equation}
\phi_{1}(r=R,\theta=\pi)-\phi_{1}(r=R,\theta=0)= \frac{d}{\lambda f} \left( R \pi \right),
\end{equation}
where $R$  is the radial size of the beam. The maximum OAM that can be sorted correctly is therefore $\ell_{max} \hbar$, where
\begin{equation}
\ell_{max} \ll \frac{d R}{2 \lambda f}.
\end{equation}

\section{The electrostatic sorter}  
Whereas in light optics a phase distribution can be imparted by a local variation in refractive index and the relative optical distance, for electrons the phase is influenced by electromagnetic potentials.  According to the standard high energy or phase object approximation \cite{Pozzi:2016}, the electron optical phase shift $\varphi (x,y)$ is given by the equation
\begin{equation}  \label{fm2p3}   
\varphi(x,y)  = {{\pi }\over {\lambda E}} \int_{-\infty}^\infty V(x,y,z) dz -{{e}\over \hbar} \int_{-\infty}^\infty {A_z}(x,y,z) dz,       
\end{equation}
where $e$ is the absolute value of the electron charge, $\hbar$ is the reduced Planck constant, $\lambda$ is the relativistically corrected de Broglie electron wavelength, $ e E$ is the relativistically corrected electron energy, $V(x,y,z)$ is the electrostatic potential and ${A_z}(x,y,z)$ is the $z$~component of the magnetic vector potential.

McMorran and co-workers \cite{Mcmorran:2017} have shown that the function of sorter~1 can be performed by a charged tip, by considering the limiting case of the analytical expression found in a previous study \cite{Matteucci:1992}, while that of sorter~2 can be realised by a periodic array of positive and negative electrodes. An alternative and quicker approach, which is based on the relationship between the two-dimensional phase shift and the projected charge \cite{Beleggia:2011}, is to take the Laplacian of the phase of sorter~1, which corresponds to the projected charge density. The result of a numerical calculation clearly demonstrates that the discontinuity line can be described by a line of projected constant charge density.

\subsection{ The unwrapper or sorter~1}
The starting point of the following considerations is not the basic equation derived previously \cite{Matteucci:1992} but a more recent and versatile equation derived for interpreting caustic phenomena associated with two oppositely biased tips \cite{Tavabi:2015}. For a charged line segment with a constant charge density $K$, lying between $(0,-a,0)$ and  $(0,0,0)$, with a compensating charge at the arbitrary far-away position $(x_D,y_D,0)$, the electrostatic potential in the $z~=~0$ plane is given by the expression
\begin{eqnarray} \label{pot}
V(x,y,0)&=&C_V\left[ \ln
 \left(\frac{\sqrt{(a+y)^2+x^2}+a+y}{\sqrt{x^2+y^2}+y}\right)-\right. \nonumber \\ &&\left. \frac{a}{\sqrt{(x+x_D)^2+(y+y_D)^2}}\right],
\end{eqnarray}
where
\begin{equation} \label{defcV}
C_V=\frac{  K}{4 \pi  \epsilon _0}.
\end{equation}
The associated phase shift takes the form
\begin{eqnarray} \label{phase}
\varphi(x,y)&=&C_E~C_V \left[ -(a+y) \ln \left((a+y)^2+x^2\right)/ \delta^2 \right)+ \nonumber \\ &&\left. y \ln \left(\left(x^2+y^2\right)/\delta^2 \right)+ 2 a+2 x \arctan \left(\frac{y}{x}\right)- \right. \nonumber \\ &&\left. 2 x \arctan \left(\frac{a+y}{x}\right)+
a \ln \left(((x+x_D)^2+  \right. \right. \nonumber \\ &&\left. \left. \left. (y+y_D)^2\right)/\delta^2 \right) \right],
\end{eqnarray}
where $C_E={{\pi }/ {\lambda E}} $ is an interaction constant that depends only on the energy of the electron beam and takes a value of $6.53\times10^6\, {\rm rad}\ {\rm V}^{-1} {\rm m}^{-1}$ for 300~kV electrons and $\delta$ is a scale factor with the dimension of length.
Any combination of line charges of constant charge density can be described by rotating and displacing the above expressions. 

\begin{figure}[ht] \centering 
\includegraphics[width=12cm]{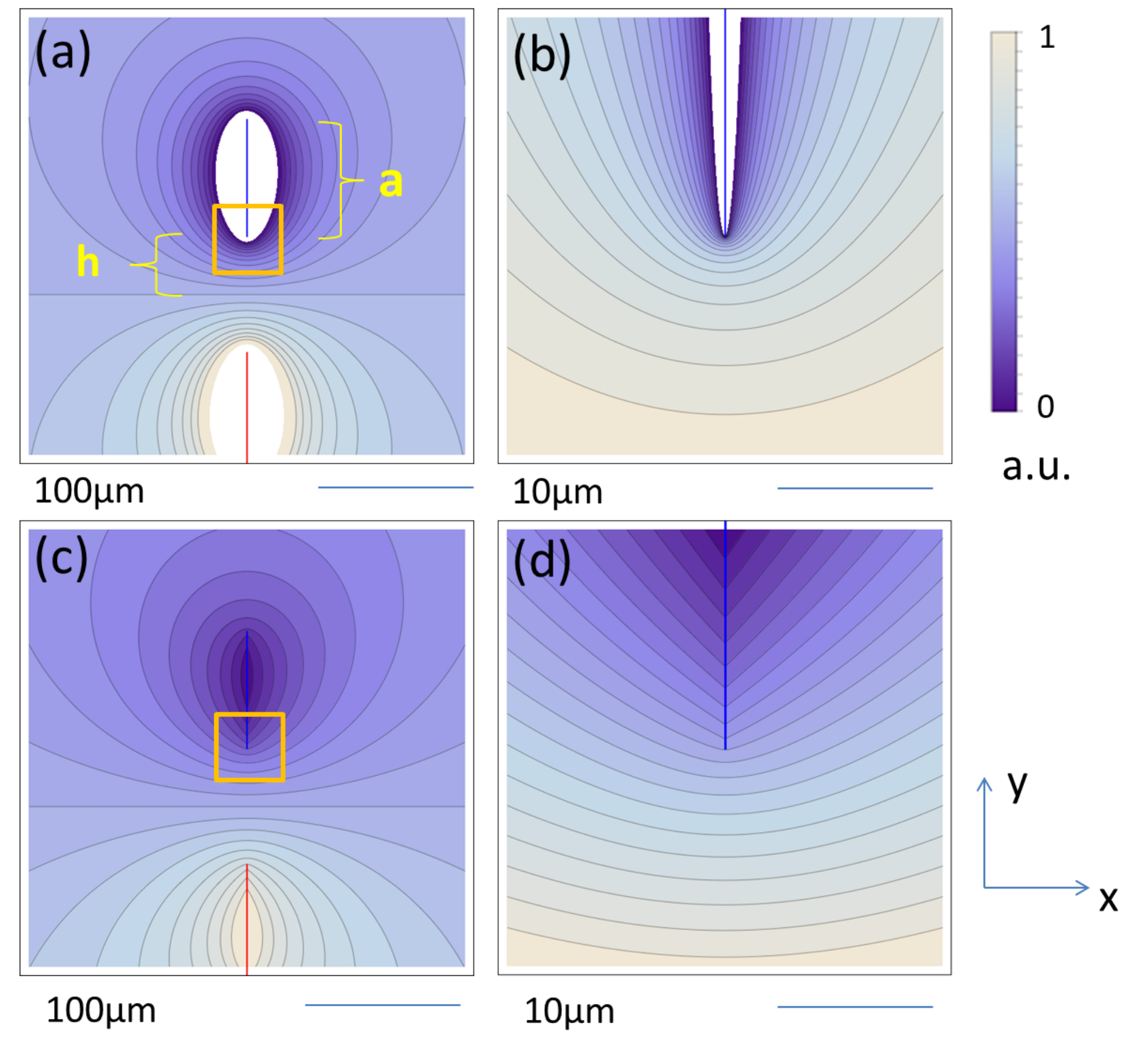} 
\caption{ (a,~b) Contour maps of the electrostatic potential (in the plane $z~=~0$) for a negatively charged line (blue) of length $a$ in front of a conducting plane at distance $h$. (c,~d) Contour maps of the corresponding phase shift. The images in (a) and (c) have a field of view of $300~\mu$m $\times$ $300~\mu$m and also show the positive image line charge (red) below the conducting plane. The centre of the image, which is marked by a square, corresponds to the tip of the negatively charged line, which is used as sorter~1. (b) and (d) show the region in the square and have a field of view of $30~\mu$m $\times$ $30~\mu$m.}
\label{sorterNJPfig1e2pot}  
\end{figure}

According to electrostatic image theory, the case of a line charge segment in front of a conducting plane is equivalent to that of two oppositely charged line segments that are placed symmetrically with respect to the mirror plane in front of it. If $h$ is the distance between one tip and the mirror plane and $a$ is the length of the line segment oriented perpendicular to the mirror plane itself, Figs~\ref{sorterNJPfig1e2pot}(a) and \ref{sorterNJPfig1e2pot}(b) show contour maps of the electrostatic potential in the $z~=~0$ plane for $a~=~80~ \mu$m and  $h~=~40~ \mu$m, calculated using Eq.~\ref{pot}. The positions of the charged lines are indicated by a blue segment (negative) and a red segment (positive). Figures~\ref{sorterNJPfig1e2pot}(c) and \ref{sorterNJPfig1e2pot}(d) show contour maps of the electron optical phase shift calculated using Eq.~\ref{phase}. The field of view in (a) and (c) is $300~\mu$m $\times$ $300~\mu$m, while in (b) and (d) it is $30~\mu$m $\times$ $30~\mu$m.

\subsection{ The corrector or sorter~2}
It is tantalising that the phase shifting effect of sorter~2 corresponds closely to that of an array of alternating reverse-biased \emph{p-n} junctions in a half-plane. The latter scenario was investigated more than 30~years ago by members of the electron holography group in Bologna \cite{Frabboni:1987, Frabboni:1985}, who derived analytical solutions for both the field \cite{Beleggia:2000a} and the phase shift \cite{Fazzini:2005}.

For the present purpose, only the phase shift associated with the fringing field protruding from the edge of the sorter 2 element into vacuum is required. In the half-space $x~>~0$, for the Fourier component $\exp( i \pi n  y/ w)$ the phase shift takes the form
\begin{equation} \label{phasearray}
\varphi_n(x,y)=2 C_E V_R \gamma_n \exp\left( i y \frac{ \pi n } {w}\right)  \frac{\exp\left( -x \vert \frac{ \pi n } {w}\vert \right) }{\sqrt{2 \vert \frac{ \pi n } {w}\vert}},
\end{equation}
where $w$ is the stripe width (with total periodicity 2$w$),  $V_R$ is the applied potential, $\gamma_n$ is the Fourier coefficient of the Fourier series expansion of the potential across the stripes and the factor 2 accounts for the contribution from the upper and lower half-spaces \cite{Beleggia:2000a,Fazzini:2005}. 

\begin{figure}[ht] \centering 
\includegraphics[width=12cm]{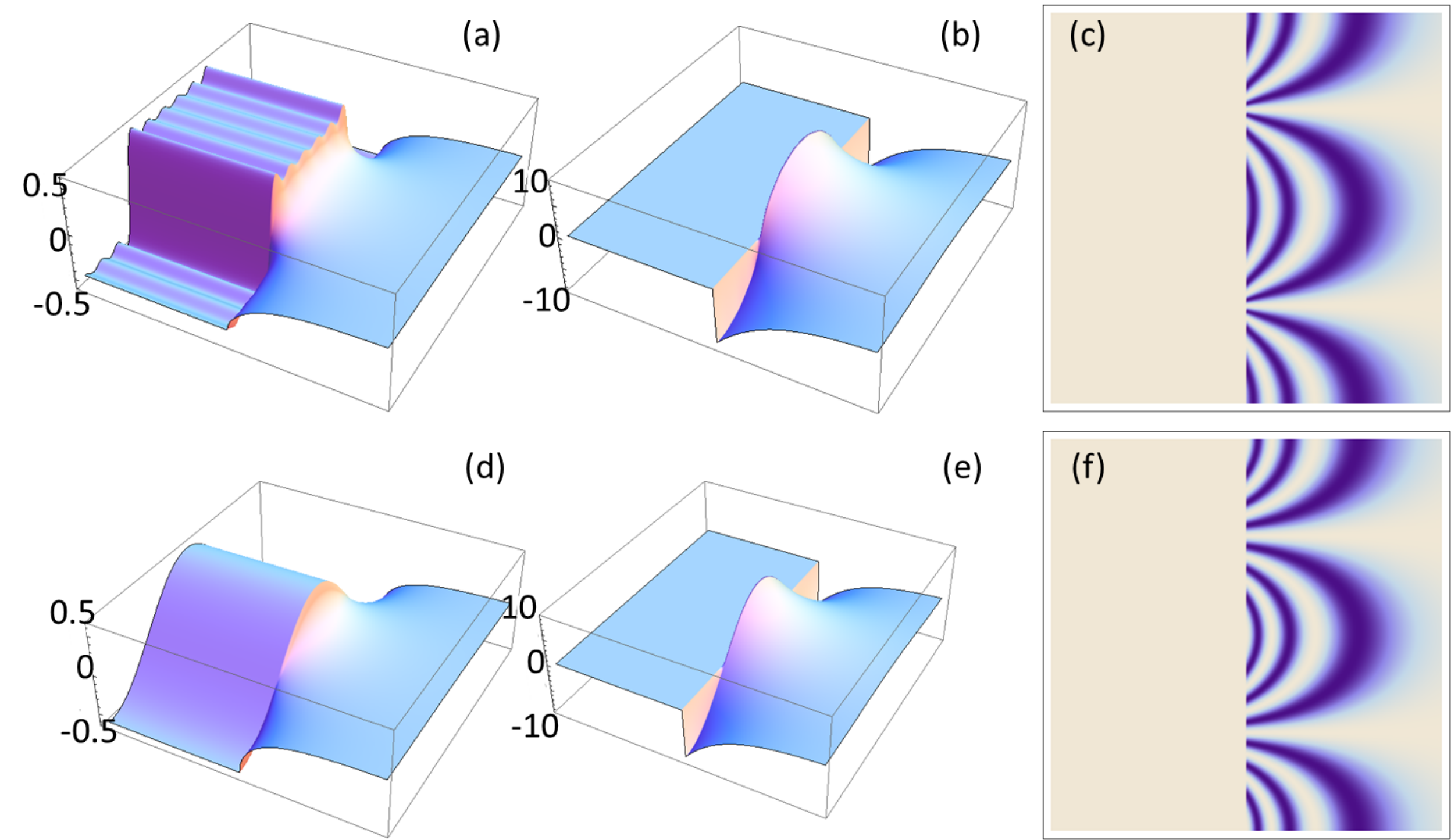} 
\caption{(a,~b)~$z=0$ sections of three-dimensional (a) potential and (b) phase distributions for stripes at alternating potentials. The maximum Fourier component of the phase corresponds to $n~=~11$. The corresponding cosine phase distribution is shown in (c). 
(d~e)~$z=0$ sections of three-dimensional (d) potential and (e) phase distributions and (f) corresponding cosine phase distribution for the first Fourier component $n~=~1$.
The sides of the square boxes and images are $20~\mu$m.}   
\label{fieldfasesorter2n5}  
\end{figure}

Figure~\ref{fieldfasesorter2n5}(a) shows the $z~=~0$ section of the three-dimensional potential distribution (in Volts), while Fig.~\ref{fieldfasesorter2n5}(b) shows the two-dimensional phase shift (in radians) in vacuum (the phase in the thick specimen is set to 0)  and Fig.~\ref{fieldfasesorter2n5}(c) shows a corresponding cosine phase contour map for a stepwise potential distribution with $V_R~=~1$V and $w~=~10~\mu$m calculated on a discrete grid of 256 $\times$ 256 points over a square of side $20 ~ \mu$m with a maximum of $n~=~11$.

Figures~\ref{fieldfasesorter2n5}(d-f) show the same calculations performed for only the first Fourier coefficient, resulting in the same functional dependence of sorter~2, Eq. (\ref{optcorr}). It is apparent that differences between the phases are much less evident than those between the potentials.

\section{Ideal \emph{vs} electrostatic sorter}  
  
For the ideal sorter, we used parameters used in experiments carried out using phase holograms made from nanofabricated SiN membranes \cite{Grillo:2017, Grillo:2014}, \emph{i.e.},  $\lambda~=~1.97$~pm (corresponding to 300~kV electrons), $f~=~1.4$~m  and  $d~=~20~\mu$m.  For these values, we obtained
\begin{equation}
\frac{d}{\lambda f}=\frac{1}{0.1378} ~ \mu{\rm m}^{-1}.
\end{equation}
As the functional form of Eq.~\ref{phase} is similar to that for sorter~1 (Eq.~\ref{optsort}), a correspondence between the ideal and electrostatic sorter~1 can be obtained if the following relationship holds:
\begin{equation} \label{relazione}
\frac{d}{2 \lambda f}=C_E C_V.
\end{equation}
According to Eqs~\ref{relazione} and \ref{defcV}, recalling that $C_E=6.53\times10^6\, {\rm rad}\ {\rm V}^{-1} {\rm m}^{-1}$ for 300~kV electrons, 
\begin{equation}
K=61.85  ~{\rm pC  ~m^{-1}}.
\end{equation}
Given the freedom of choice for the parameter $b$, in order to compare the ideal and electrostatic sorter~1 it is useful to subtract the linear contribution over the  field of view of $30~\mu$m in both cases, \emph{i.e.}, to use de-linearized phases. This procedure is also useful for reducing artefacts at the edges of the images during calculations.

\begin{figure}[ht] \centering 
\includegraphics[width=12cm]{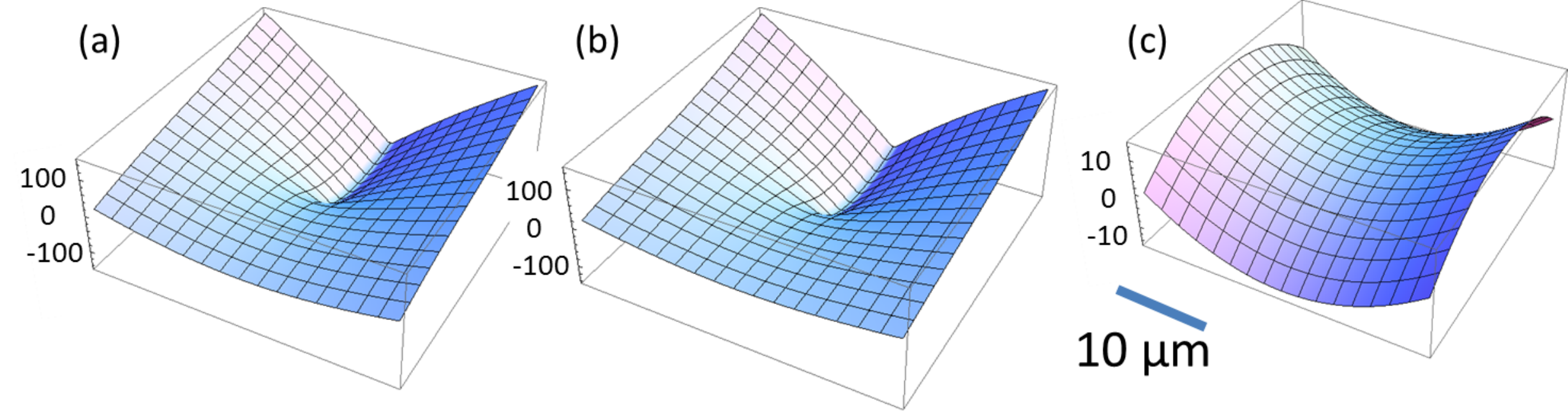} 
\caption{Three-dimensional representations of the de-linearized phase shift of (a) an ideal sorter~1, (b)~its electrostatic counterpart and (c)~their difference for  $a~=~80~\mu$m and $h~=~40~\mu$m. The side of the square box  is $30~\mu$m.}   
\label{sorterNJPFig5}  
\end{figure}

Figure~\ref{sorterNJPFig5}(a) shows the de-linearized phase for an ideal sorter~1 (a) according to Eq.~\ref{optsort}, Fig.~~\ref{sorterNJPFig5}(b) shows the de-linearized phase of its electrostatic counterpart according to Eq.~\ref{phase} and their difference is shown in Fig.~\ref{sorterNJPFig5}(c). The corresponding charged line has a length of $a~=~80~\mu$m and is placed at a distance of $h~=~40~\mu$m from a conducting plane.
If $a$ and $h$ are multiplied by a factor 1000, then the two phases are practically coincident.
It is therefore apparent that the finite length of the charged line and its distance from the conducting plane (\emph{i.e.}, the boundary condition) affect the phase shift strongly. The saddle shape of the difference image is strongly reminiscent of an additional astigmatic contribution to the phase. We demonstrate below its effect on the operation of the sorter and we propose an approach for its circumvention.

\section{Simulations} 
\subsection{The ideal sorter}

In order to compare the performance of the ideal and electrostatic sorter, we assumed that the illuminating electron beam $\psi_{ill}(x,y)$ that impinges on sorter~1 takes the form of a superposition of Laguerre-Gaussian beams, whose wavefunction at their waist is given by the expression \cite{Lavery:2011} 
\begin{eqnarray} 
\psi^{LG}_{\ell,p}(x,y,w_0)=\frac{1}{\sqrt{\pi } w_0} 2^{\frac{\left| \ell \right| }{2}+\frac{1}{2}} e^{-\frac{x^2+y^2}{w_0^2}+i (2p+|\ell|+1) \,\arctan{(y/x)}}
\nonumber \\ 
\left(\frac{\sqrt{x^2+y^2}}{w_0}\right)^{\left| \ell \right| } \sqrt{\frac{p!}{(\left| \ell \right| +p)!}} L_p^{\left| \ell \right| }\left(\frac{2 \left(x^2+y^2\right)}{w_0^2}\right),
\end{eqnarray}
where $w_0$ is the waist size and  $ L_p^{\left|  \ell \right| } (x)$ is the associated Laguerre polynomial for mode indices $\ell$ and $p$.

In particular, we chose a superposition of two Laguerre-Gaussian beams with $w_0~=~7~\mu$m, $\ell~=~\pm1$ and $p~=~0$, which can be written in the form
\begin{equation} \label{psill}
\psi_{ill}(x,y)=\psi^{LG}_{1,0}(x,y,w_0)+\psi^{LG}_{-1,0}(x,y,w_0).
\end{equation}
The wavefunction immediately after sorter~1 is given by the product of the illumination (Eq.~\ref{psill}) and the transmission function of sorter~1 plus the thin lens of focal length $f$ (Eq.~\ref{S1piuf}). Propagation in field-free space between sorter~1 and sorter~2 can be calculated by using a simulation program that is based on the Fast Fourier Transform evaluation of the  Kirchhoff-Fresnel diffraction integral \cite{Beleggia:2014,Tavabi:2015}. Artefacts resulting from periodic continuation can be minimized by using the same square region of side $30 \mu$m with 512 $\times$ 512 pixels and removing the linear phase gradient, \emph{i.e.}, by using the de-linearized phase.

\begin{figure}[ht] \centering 
\includegraphics[width=12cm]{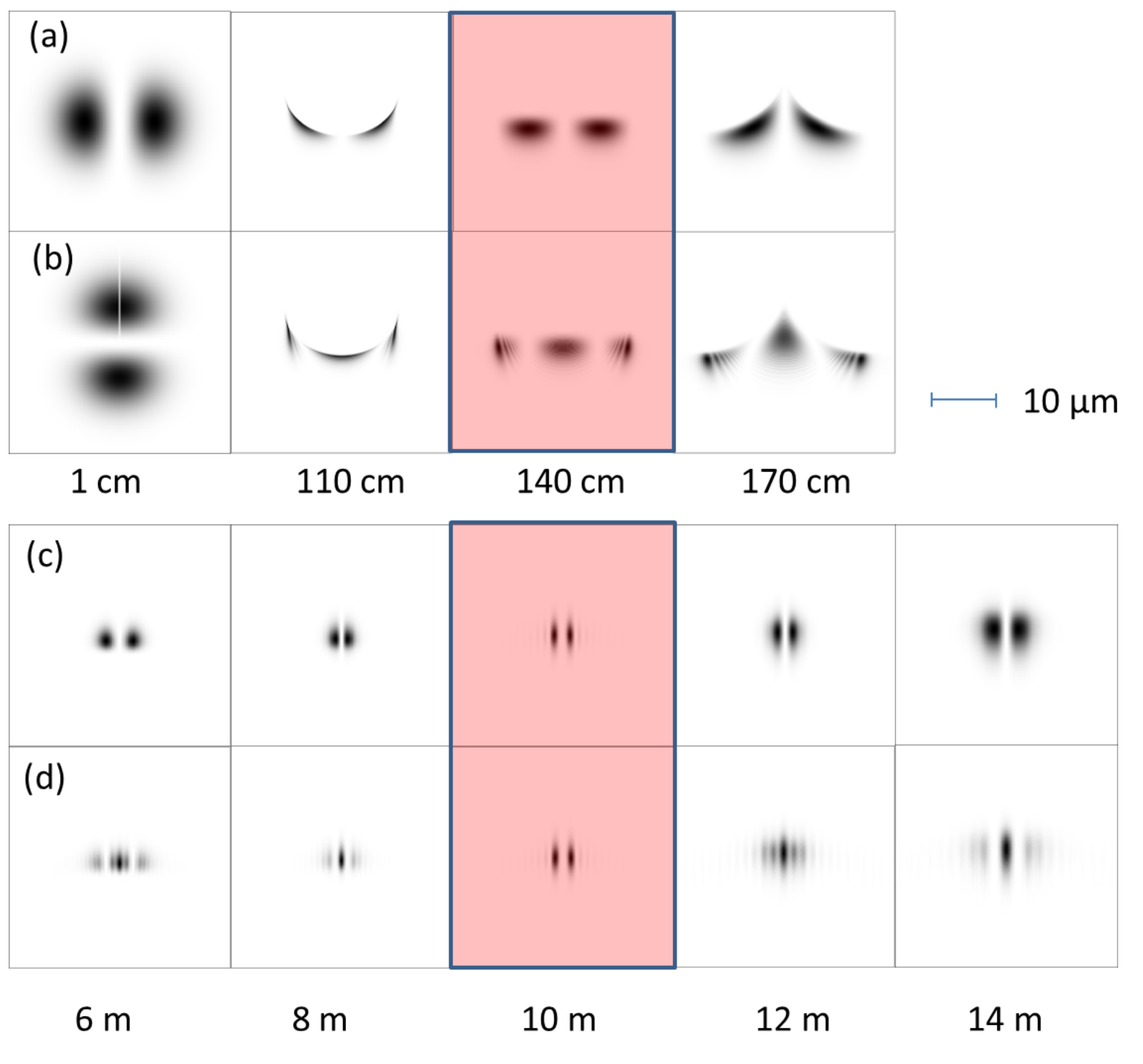} 
\caption{ (a,~b)~Fresnel defocus images after the ideal unwrapper or sorter 1 at distances (from left to right) of 1~cm (nearly in focus), 110~cm, 140~cm  (corresponding to a Fraunhofer image) and 170~cm. The probes are superimposed beams given by Eq.~\ref{psill}, with the zeros set to be (a) parallel and (b) perpendicular to the discontinuity. (c,~d)~The same beams as in (a) and (b) propagated after the ideal sorter~2 or corrector at defocus values of 6~m, 8~m, 10~m (corresponding to the camera length of the Fraunhofer diffraction pattern), 12~m and 14~m.}   
\label{njpsorter7}  
\end{figure}

Figure~\ref{njpsorter7}(a) shows Fresnel defocus images calculated for several distances from the illuminating beam given by Eq.~\ref{psill}. The first distance of 1~cm is very close to focus and and therefore shows the intensity distribution of the illuminating beam. The second distance is 110~cm. The third distance is 140~cm, corresponds to the distance between sorter~1 and sorter~2 (equal to the focal distance $f$ of the lens) and is the Fraunhofer or diffraction image of sorter~1, which will subsequently be processed by sorter~2. At the fourth distance of 170~cm, the image is detected in the absence of sorter~1. The images at 140 and 170~cm correspond to overfocused and underfocused images of charged tips \cite{Beleggia:2014}.
The simulations in Fig.~\ref{njpsorter7}(d) also indicate that reliable OAM sorting only works for an exact defocus condition, since an unforeseen peak at $\ell~=~0$ appears at an out of focus condition.
By changing the $x$ and $y$ coordinates in the illumination, the same set of images is obtained in Fig.~\ref{njpsorter7}(b). In particular, the bright line at 1~cm corresponds to a slightly defocused image of the discontinuity line of sorter~1. 

We now consider the wavefunction at 140~cm as the illuminating beam for sorter~2, described by the transmission function Eq.~\ref{S2piuf}. We add another thin lens of focal length $F~=~10$~m, in order to obtain images at a suitable magnification. Figures~\ref{njpsorter7}(c) and \ref{njpsorter7}(d) show results corresponding to Figs~\ref{njpsorter7}(a) and \ref{njpsorter7}(b), respectively, calculated for defocus values of 6~m, 8~m, 10~m (corresponding to the camera length of the Fraunhofer diffraction pattern), 12~m and 14~m.
The two spots are well separated in the diffraction plane, as expected.
 
In both cases, slight defocus introduces an apparent distortion of the OAM spectrum, which is different between Fig.~\ref{njpsorter7}(c) and Fig.~\ref{njpsorter7}(d).  For the series in Fig.~\ref{njpsorter7}(d), it appears that a component $\ell~=~0$ is dominant in the spectrum.

\subsection{The electrostatic sorter}
\begin{figure}[b] \centering 
\includegraphics[width=12cm]{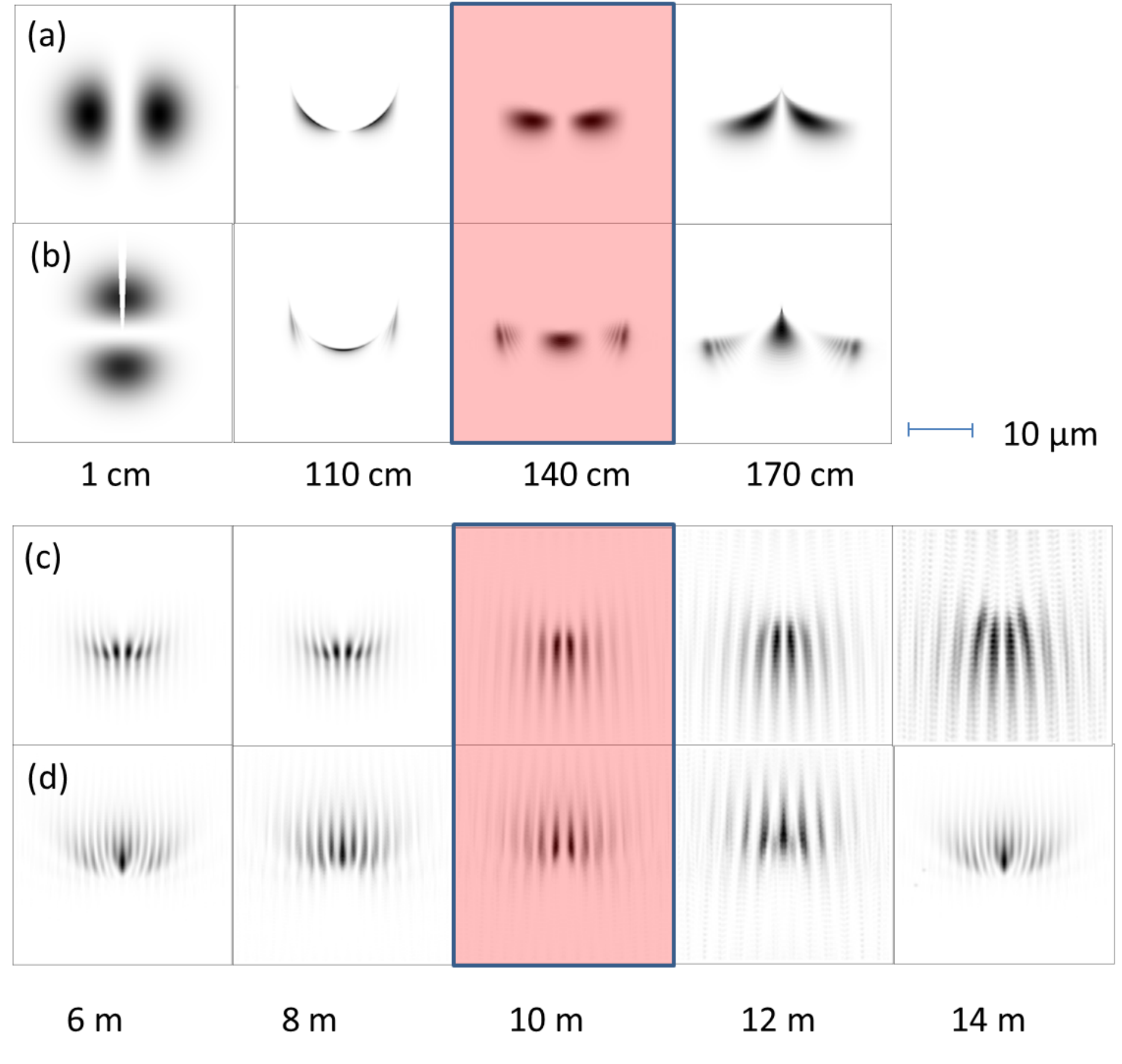} 
\caption{ (a,~b)~Fresnel defocus images after the electrostatic unwrapper or sorter 1 at distances (from left to right) of 1~cm (nearly in focus), 110~cm, 140~cm  (corresponding to a Fraunhofer image) and 170~cm. The probes are superimposed beams given by Eq.~\ref{psill}, with the zeroes set to be (a) parallel and (b) perpendicular to the discontinuity. (c,~d)~The same set of beams in (a) and (b) propagated after the electrostatic sorter~2 or corrector at defocus values of 6~m, 8~m, 10~m (corresponding to the camera length of the Fraunhofer diffraction pattern), 12~m and 14~m.}   
\label{njpsorter11}  
\end{figure} 

In this section, the above calculations are repeated for the electrostatic sorter by using the corresponding phase shifts in place of  $\phi_1$ and $\phi_2$. The images are calculated using the same parameters as in the previous section and are displayed in the same manner.

For sorter~1, with a fixed charge density, we used a potential of 5~V. For sorter~2, we used the first term in the Fourier series expansion (\emph{i.e.}, we considered the ideal behaviour) and set the edge to be at a distance of - 4~$\mu$m from the centre. The applied potential was chosen to be $V_R~=~-12.3$ V. In Fig.~\ref{njpsorter11}(a), the image of the tip falls in a region of negligible intensity, whereas in Fig.~\ref{njpsorter11}(b) it is visible. The influence of astigmatism introduced by the finite length of the line charge and the boundary condition can be seen to almost mask the effect of the electrostatic sorter.
A comparison of the ideal spectrum with images (c,~d) at 10~m shows that, instead of the 2~lines that are expected from the theoretical spectrum, a set of parallel lines is formed instead. A reliable OAM spectrum of the beam therefore cannot be obtained.

\begin{figure}[b] \centering 
\includegraphics[width=8cm]{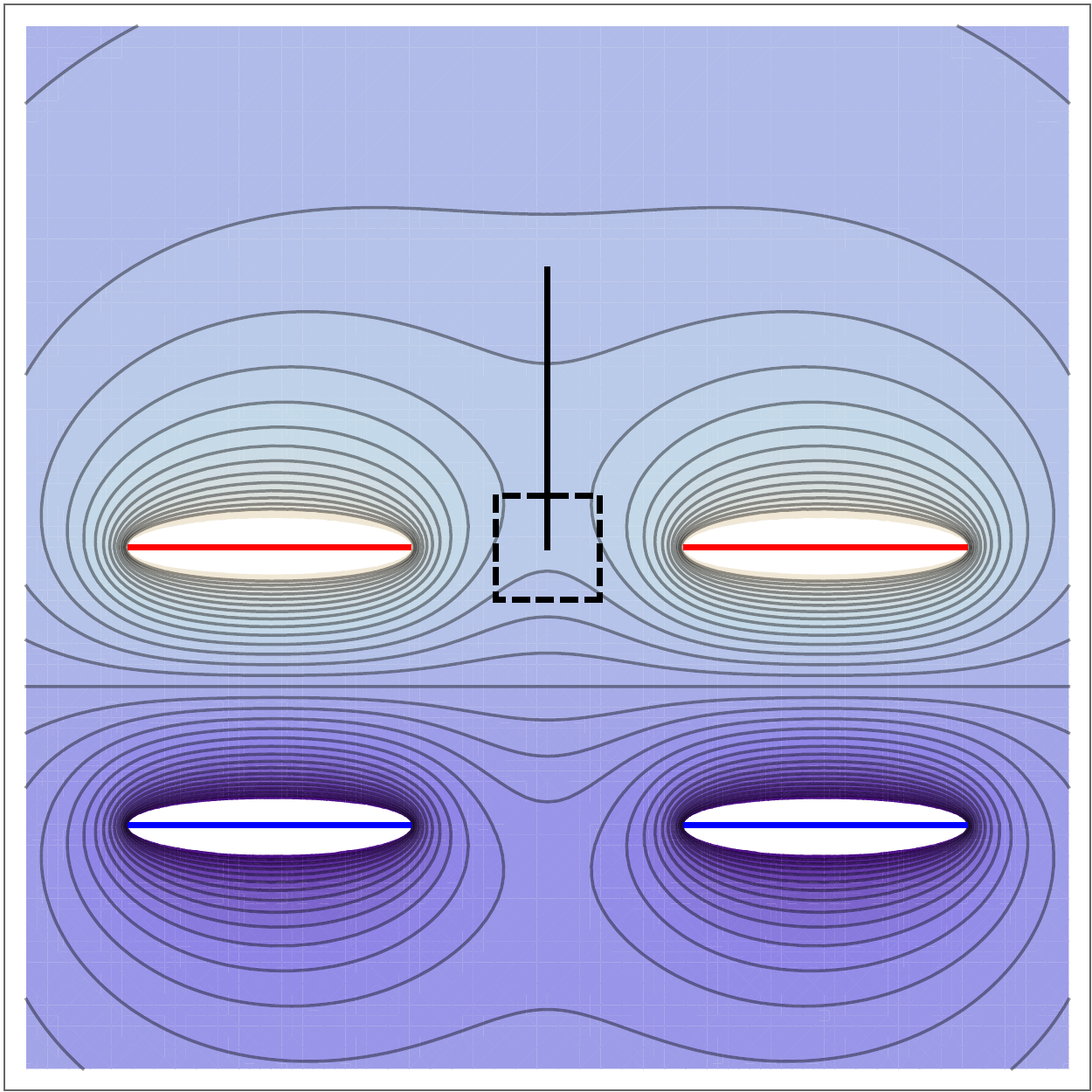} 
\caption{Contour map of the electrostatic potential in the plane $z~=~0$ for two charged lines (red) parallel to a conducting plane, together with their mirrored images (blue). The potential of sorter~1 is not applied, but its position is indicated by the dark line. The dashed square (of side $30~\mu$m) indicates the region used in the calculations. The side of the image is $300~\mu$m.}   
\label{astcorrpot}  
\end{figure} 
\section{Astigmatism reduction}
In order to compensate, at least partially, for the astigmatism of sorter~1, we investigated a more elaborate configuration of charged lines,  as shown in Fig.~\ref{astcorrpot}, by including two more charged lines (red) of equal charge density parallel to the conducting plane, with their mirrored images (blue) on the opposite side. These charged lines have the same length and charge density as that of sorter~1. The potential of sorter~1 is not included in the figure, but its position is indicated by the dark line. The dashed box marks the region used in the calculations.
\begin{figure}[ht] \centering 
\includegraphics[width=8cm]{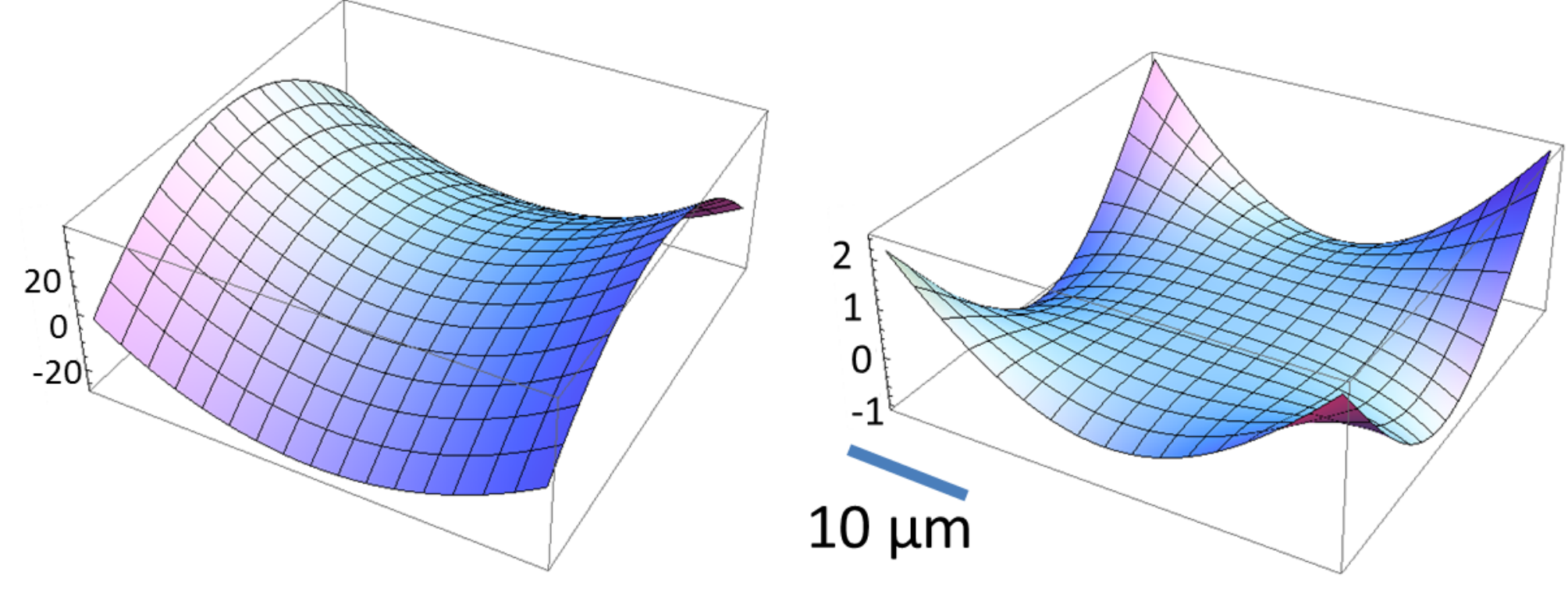} 
\caption{(a)~Three-dimensional representation of the de-linearized phase of two  charged lines parallel to a conducting plane in the square region marked in Fig.~\ref{astcorrpot}. (b)~Three-dimensional representation of the difference between the de-linearized phase of two charged lines parallel to a conducting plane (a)  multiplied by -0.546 and  the astigmatic term in the phase of sorter~1 (Fig.~\ref{sorterNJPFig5}(c)).}   
\label{astig3d}  
\end{figure} 

Figure~\ref{astig3d}(a) shows a three-dimensional representation of the de-linearized phase of two charged lines parallel to a conducting plane in the region of the tip of sorter~1. A comparison with Fig.~\ref{sorterNJPFig5}(c) suggests that, if the charge density is reduced by a factor of 0.546 and its sign is changed, then the phase of the two parallel line charges almost compensates for the astigmatic term of sorter~1, as shown in Fig.~\ref{astig3d}(b).
\begin{figure}[t] \centering 
\includegraphics[width=12cm]{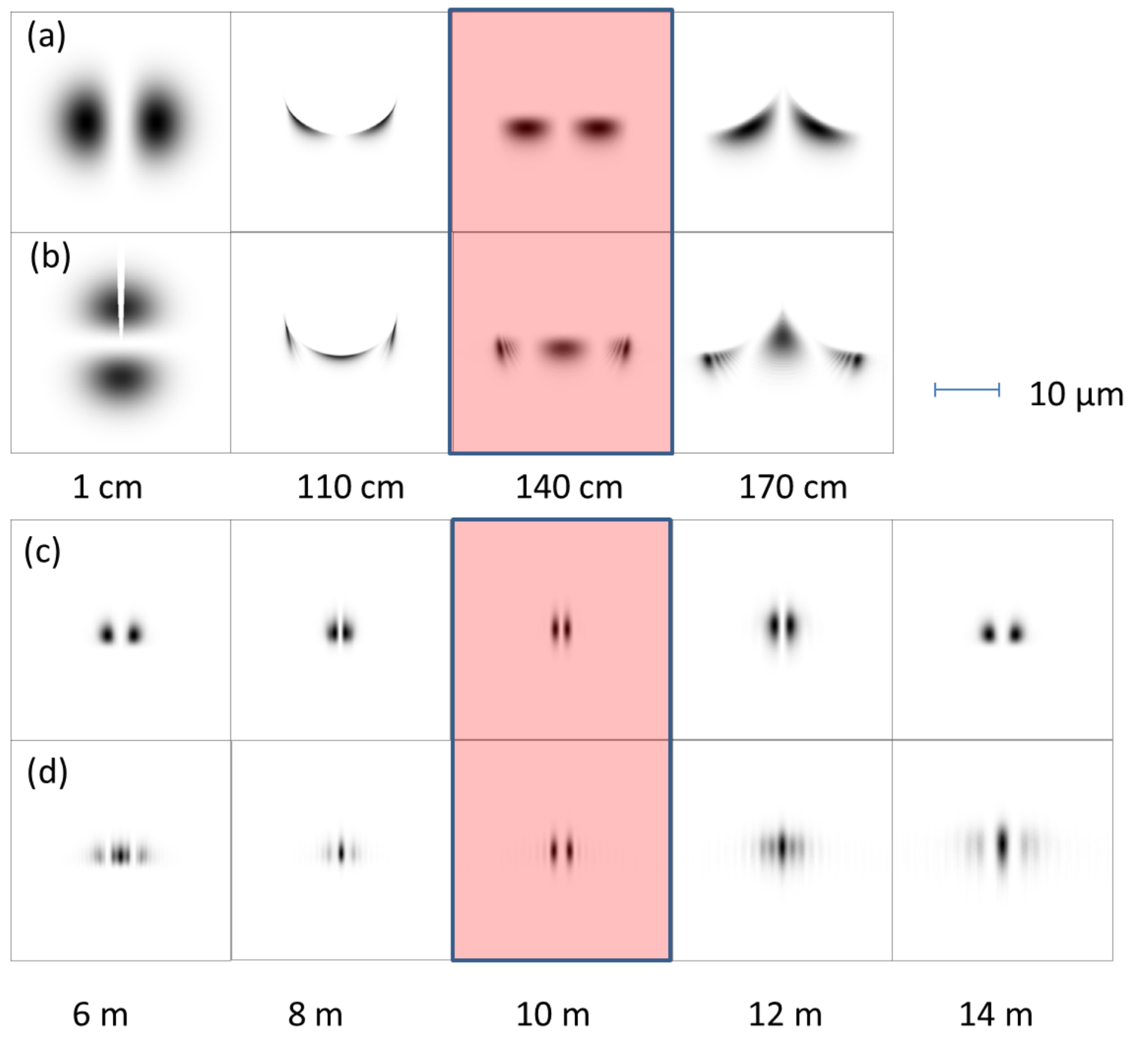} 
\caption{ (a,~b)~Fresnel defocus images for the new configuration of the electrostatic unwrapper or sorter~1 at distances, from left to right, of 1~cm (nearly in focus), 110~cm, 140~cm  (corresponding to a Fraunhofer image) and 170~cm. The probes are superimposed beams given by Eq.~\ref{psill}, with the zeros set to be (a) parallel and (b) perpendicular to the discontinuity. (c,~d) The same set of beams as in (a) and (b) propagated after the electrostatic sorter~2 or corrector at defocus values of 6~m, 8~m, 10~m (corresponding to the camera length of the Fraunhofer diffraction pattern), 12~m and 14~m.}   
\label{njpsorter18}  
\end{figure} 
We therefore replaced sorter~1 with the new compensated configuration made by three line charges (and three mirrored images). The results are shown in Fig.~\ref{njpsorter18}. The quality of the results is greatly improved and can be compared favourably to the ideal sorter.

\section{Discussion and conclusions}
In this work we analyzed by simulations the realistic conditions of the parameters and imperfection for the two phase elements producing an electrostatic OAM sorter for electrons. 
While the ideal phase elements  comprise for the first element a long charged needle and for the second a series of alternating potentials, particular care must be used for the actual device. 
Simulations performed for the electrostatic sorter show that the electrostatic potentials that are required for either sorter~1 (a few Volts) or sorter~2 (a few tens of Volts) are not prohibitive and are realisable experimentally, as demonstrated by our experiments on biased tips \cite{Beleggia:2014,Tavabi:2015} and reverse-biased \emph{p-n} junctions \cite{Frabboni:1987}.
Results also indicate that the detailed shape of electrodes in sorter 2 does not affect the actual potential, provided a sufficient distance of the beam from  the electrodes.
The most important and detrimental effect results from the astigmatic term in sorter~1, when the needle has finite length, which makes secondary diffraction maxima more intense in the final observation plane and increases cross-talk between neighboring OAM states. This effect results from the finite length of the line charge, as well as from its mirror image and the boundary condition. Although it may be possible to use a correcting quadrupole after sorter~1, or to use the electron microscope astigmatism correctors, we have shown that the addition of two suitably charged lines perpendicular to sorter~1 and parallel to the conducting plane can lead to satisfactory compensation of such astigmatism.

\ack
The authors acknowledge the support of the  European Union’s Horizon 2020 Research and Innovation Programme under Grant Agreement No 766970 Q-SORT (H2020-FETOPEN-1-2016-2017).
 E.K. acknowledges the support of a Canada Research Chair and Ontario's Early Researcher Award. This project has received funding from the European Union's Horizon 2020 Research and Innovation Programme under grant agreement No. 823717-ESTEEM3.

\section*{References}


\begin{thebibliography}{10}

\bibitem{Noether:1918}
Emmy Noether.
\newblock Invariante Variationsprobleme.
\newblock {\em Nachr. D. K{\"o}nig. Gesellsch. D. Wiss. Zu G{\"o}ttingen,
  Math-Phys. Klasse}, pages 235--257, 1918.

\bibitem{Neuenschwander:2017}
Dwight~E Neuenschwander.
\newblock {\em Emmy Noether's wonderful theorem}.
\newblock JHU Press, 2017.

\bibitem{Bliokh:2007}
Konstantin~Y Bliokh, Yury~P Bliokh, Sergey Savel'ev, and Franco Nori.
\newblock Semiclassical dynamics of electron wave packet states with phase
  vortices.
\newblock {\em Physical Review Letters}, 99(19):190404, 2007.

\bibitem{Uchida:2010}
Masaya Uchida and Akira Tonomura.
\newblock Generation of electron beams carrying orbital angular momentum.
\newblock {\em Nature}, 464(7289):737--739, 2010.

\bibitem{Verbeeck:2010}
Johan Verbeeck, He~Tian, and Peter Schattschneider.
\newblock Production and application of electron vortex beams.
\newblock {\em Nature}, 467(7313):301--304, 2010.

\bibitem{Mcmorran:2011}
Benjamin~J McMorran, Amit Agrawal, Ian~M Anderson, Andrew~A Herzing, Henri~J
  Lezec, Jabez~J McClelland, and John Unguris.
\newblock Electron vortex beams with high quanta of orbital angular momentum.
\newblock {\em Science}, 331(6014):192--195, 2011.

\bibitem{Harris:2015}
J{\'e}r{\'e}mie Harris, Vincenzo Grillo, Erfan Mafakheri, Gian~Carlo Gazzadi,
  Stefano Frabboni, Robert~W Boyd, and Ebrahim Karimi.
\newblock Structured quantum waves.
\newblock {\em Nature Physics}, 11(8):629--634, 2015.

\bibitem{Mafakheri:2017}
Erfan Mafakheri, Amir~H Tavabi, Peng-Han Lu, Roberto Balboni, Federico Venturi,
  Claudia Menozzi, Gian~Carlo Gazzadi, Stefano Frabboni, Alicia Sit, Rafal~E
  Dunin-Borkowski, Ebrahim Karimi, and Vincenzo Grillo.
\newblock Realization of electron vortices with large orbital angular momentum
  using miniature holograms fabricated by electron beam lithography.
\newblock {\em Applied Physics Letters}, 110(9):093113, 2017.

\bibitem{Bliokh:2017}
Konstantin~Y Bliokh, Igor~P Ivanov, Giulio Guzzinati, Laura Clark, Ruben
  Van~Boxem, Armand B{\'e}ch{\'e}, Roeland Juchtmans, Miguel~A Alonso, Peter
  Schattschneider, Franco Nori, and Jo~Verbeeck.
\newblock Theory and applications of free-electron vortex states.
\newblock {\em Physics Reports}, 690:1--70, 2017.

\bibitem{Larocque:2018}
Hugo Larocque, Ido Kaminer, Vincenzo Grillo, Gerd Leuchs, Miles~J. Padgett,
  Robert~W. Boyd, Mordechai Segev, and Ebrahim Karimi.
\newblock `twisted' electrons.
\newblock {\em Contemporary Physics}, 59(2):126--144, 2018.

\bibitem{Grillo:2017}
Vincenzo Grillo, Amir~H Tavabi, Federico Venturi, Hugo Larocque, Roberto
  Balboni, Gian~Carlo Gazzadi, Stefano Frabboni, Peng-Han Lu, Erfan Mafakheri,
  Fr{\'e}d{\'e}ric Bouchard, Rafal~E Dunin-Borkowski, Robert~W Boyd, Martin~PJ
  Lavery, Miles~J Padgett, and Ebrahim Karimi.
\newblock Measuring the orbital angular momentum spectrum of an electron beam.
\newblock {\em Nature Communications}, 8, 2017.

\bibitem{Mcmorran:2017}
Benjamin~J McMorran, Tyler~R Harvey, and Martin~PJ Lavery.
\newblock Efficient sorting of free electron orbital angular momentum.
\newblock {\em New Journal of Physics}, 19(2):023053, 2017.

\bibitem{Berkhout:2010}
Gregorius~CG Berkhout, Martin~PJ Lavery, Johannes Courtial, Marco~W
  Beijersbergen, and Miles~J Padgett.
\newblock Efficient sorting of orbital angular momentum states of light.
\newblock {\em Physical Review Letters}, 105(15):153601, 2010.

\bibitem{Lavery:2012}
Martin~PJ Lavery, David~J Robertson, Gregorius~CG Berkhout, Gordon~D Love,
  Miles~J Padgett, and Johannes Courtial.
\newblock Refractive elements for the measurement of the orbital angular
  momentum of a single photon.
\newblock {\em Optics Express}, 20(3):2110--2115, 2012.

\bibitem{Bryngdahl:1974a}
Olof Bryngdahl.
\newblock Optical map transformations.
\newblock {\em Optics Communications}, 10(2):164--168, 1974.

\bibitem{Bryngdahl:1974b}
Olof Bryngdahl.
\newblock Geometrical transformations in optics.
\newblock {\em JOSA}, 64(8):1092--1099, 1974.

\bibitem{Saito:1983}
Yoshiharu Saito, Shin-ichi Komatsu, and Hitoshi Ohzu.
\newblock Scale and rotation invariant real time optical correlator using
  computer generated hologram.
\newblock {\em Optics Communications}, 47(1):8--11, 1983.

\bibitem{Cederquist:1984}
Jack Cederquist and Anthony~M Tai.
\newblock Computer-generated holograms for geometric transformations.
\newblock {\em Applied Optics}, 23(18):3099--3104, 1984.

\bibitem{Hossack:1987}
WJ~Hossack, AM~Darling, and A~Dahdouh.
\newblock Coordinate transformations with multiple computer-generated optical
  elements.
\newblock {\em Journal of Modern Optics}, 34(9):1235--1250, 1987.

\bibitem{Grillo:2014}
Vincenzo Grillo, Gian Carlo~Gazzadi, Ebrahim Karimi, Erfan Mafakheri, Robert~W
  Boyd, and Stefano Frabboni.
\newblock Highly efficient electron vortex beams generated by nanofabricated
  phase holograms.
\newblock {\em Applied Physics Letters}, 104(4):043109, 2014.

\bibitem{Verbeeck:2018}
Jo~Verbeeck, Armand B{\'e}ch{\'e}, Knut M{\"u}ller-Caspary, Giulio Guzzinati,
  Minh~Anh Luong, and Martien~Den Hertog.
\newblock Demonstration of a 2x2 programmable phase plate for electrons.
\newblock {\em Ultramicroscopy}, 190:58 -- 65, 2018.

\bibitem{Haider:1998}
Maximilian Haider, Stephan Uhlemann, Eugen Schwan, Harald Rose, Bernd Kabius,
  and Knut Urban.
\newblock Electron microscopy image enhanced.
\newblock {\em Nature}, 392(6678):768, 1998.

\bibitem{Grillo:2017a}
Vincenzo Grillo, Amir~H Tavabi, Emrah Yucelen, Peng-Han Lu, Federico Venturi,
  Hugo Larocque, Lei Jin, Aleksei Savenko, Gian~Carlo Gazzadi, Roberto Balboni,
  Stefano Frabboni, Peter Tiemeijer, Rafal~E Dunin-Borkowski, and Ebrahim
  Karimi.
\newblock Towards a holographic approach to spherical aberration correction in
  scanning transmission electron microscopy.
\newblock {\em Optics Express}, 25(18):21851--21860, 2017.

\bibitem{Linck:2017}
Martin Linck, Peter~A Ercius, Jordan~S Pierce, and Benjamin~J McMorran.
\newblock Aberration corrected STEM by means of diffraction gratings.
\newblock {\em Ultramicroscopy}, 182:36--43, 2017.

\bibitem{Shiloh:2018}
Roy Shiloh, Roei Remez, Peng-Han Lu, Lei Jin, Yossi Lereah, Amir~H Tavabi,
  Rafal~E Dunin-Borkowski, and Ady Arie.
\newblock Spherical aberration correction in a scanning transmission electron
  microscope using a sculpted thin film.
\newblock {\em Ultramicroscopy}, 189:46--53, 2018.

\bibitem{Beleggia:2014}
Marco Beleggia, Takeshi Kasama, David~J Larson, Thomas~F Kelly, Rafal~E
  Dunin-Borkowski, and Giulio Pozzi.
\newblock Towards quantitative off-axis electron holographic mapping of the
  electric field around the tip of a sharp biased metallic needle.
\newblock {\em Journal of Applied Physics}, 116(2):024305, 2014.

\bibitem{Frabboni:1987}
Stefano Frabboni, Giorgio Matteucci, and Giulio Pozzi.
\newblock Observation of electrostatic fields by electron holography: The case
  of reverse-biased p-n junctions.
\newblock {\em Ultramicroscopy}, 23(1):29--37, 1987.

\bibitem{Beleggia:2000a}
Marco Beleggia, Raffaella Capelli, and Giulio Pozzi.
\newblock A model for the interpretation of holographic and lorentz images of
  tilted reverse-biased p-n junctions in a finite specimen.
\newblock {\em Philosophical Magazine B: Physics of Condensed Matter;
  Statistical Mechanics, Electronic, Optical and Magnetic Properties},
  80(5):1071--1082, 2000.

\bibitem{Pozzi:2016}
Giulio Pozzi.
\newblock {Particles and waves in electron optics and microscopy}.
\newblock In Peter~W Hawkes, editor, {\em Advances in Imaging and Electron
  Physics}, volume 194. Elsevier Academic Press, New York, NY, 2016.

\bibitem{Matteucci:1992}
Giorgio Matteucci, Gian~Franco Missiroli, Michele Muccini, and Giulio Pozzi.
\newblock Electron holography in the study of the electrostatic fields: the
  case of charged microtips.
\newblock {\em Ultramicroscopy}, 45(1):77--83, 8 1992.

\bibitem{Beleggia:2011}
Marco Beleggia, Takeshi Kasama, Rafal~E Dunin-Borkowski, Stephan Hofmann, and
  Giulio Pozzi.
\newblock Direct measurement of the charge distribution along a biased carbon
  nanotube bundle using electron holography.
\newblock {\em Applied Physics Letters}, 98(24):243101, 2011.

\bibitem{Tavabi:2015}
Amir~Hossein Tavabi, Vadim Migunov, Christian Dwyer, Rafal~E. Dunin-Borkowski,
  and Giulio Pozzi.
\newblock Tunable caustic phenomena in electron wavefields.
\newblock {\em Ultramicroscopy}, 157:57--64, 10 2015.

\bibitem{Frabboni:1985}
Stefano Frabboni, Giorgio Matteucci, Giulio Pozzi, and Massimo Vanzi.
\newblock Electron holographic observations of the electrostatic field
  associated with thin reverse-biased p-n junctions.
\newblock {\em Physical Review Letters}, 55(20):2196--2199, 1985.

\bibitem{Fazzini:2005}
Pier~Francesco Fazzini, Giulio Pozzi, and Marco Beleggia.
\newblock Electron optical phase-shifts by fourier methods: Analytical versus
  numerical calculations.
\newblock {\em Ultramicroscopy}, 104(3-4):193--205, 10 2005.

\bibitem{Lavery:2011}
Martin~PJ Lavery, Gregorius~CG Berkhout, Johannes Courtial, and Miles~J
  Padgett.
\newblock Measurement of the light orbital angular momentum spectrum using an
  optical geometric transformation.
\newblock {\em Journal of Optics}, 13(6):064006, 2011.

\end{thebibliography}

\end{document}